\PassOptionsToPackage{unicode}{hyperref}
\PassOptionsToPackage{hyphens}{url}
\PassOptionsToPackage{dvipsnames,svgnames,x11names}{xcolor}
\documentclass[
  10pt,
  twocolumn]{article}
\usepackage{amsmath,amssymb}
\usepackage{iftex}
\ifPDFTeX
  \usepackage[T1]{fontenc}
  \usepackage[utf8]{inputenc}
  \usepackage{textcomp} 
\else 
  \usepackage{unicode-math} 
  \defaultfontfeatures{Scale=MatchLowercase}
  \defaultfontfeatures[\rmfamily]{Ligatures=TeX,Scale=1}
\fi
\usepackage{lmodern}
\ifPDFTeX\else
\fi
\IfFileExists{upquote.sty}{\usepackage{upquote}}{}
\IfFileExists{microtype.sty}{
  \usepackage[]{microtype}
  \UseMicrotypeSet[protrusion]{basicmath} 
}{}
\makeatletter
\@ifundefined{KOMAClassName}{
  \IfFileExists{parskip.sty}{%
    \usepackage{parskip}
  }{
    \setlength{\parindent}{0pt}
    \setlength{\parskip}{6pt plus 2pt minus 1pt}}
}{
  \KOMAoptions{parskip=half}}
\makeatother
\usepackage{xcolor}
\usepackage{graphicx}
\makeatletter
\def\maxwidth{\ifdim\Gin@nat@width>\linewidth\linewidth\else\Gin@nat@width\fi}
\def\maxheight{\ifdim\Gin@nat@height>\textheight\textheight\else\Gin@nat@height\fi}
\makeatother
\setkeys{Gin}{width=\maxwidth,height=\maxheight,keepaspectratio}
\makeatletter
\def\fps@figure{htbp}
\makeatother
\NewDocumentCommand\citeproctext{}{}

\makeatletter
 \let\@cite@ofmt\@firstofone
 \def\@biblabel#1{}
 \def\@cite#1#2{{#1\if@tempswa , #2\fi}}
\makeatother
\newlength{\cslhangindent}
\newlength{\csllabelwidth}
\newenvironment{CSLReferences}[2] 
 {\begin{list}{}{%
  \setlength{\itemindent}{0pt}
  \setlength{\leftmargin}{0pt}
  \setlength{\parsep}{0pt}
  \ifodd #1
   \setlength{\leftmargin}{\cslhangindent}
   \setlength{\itemindent}{-1\cslhangindent}
  \fi
  \setlength{\itemsep}{#2\baselineskip}}}
 {\end{list}}
\usepackage{calc}

\newcommand{\CSLLeftMargin}[1]{\parbox[t]{\csllabelwidth}{\strut#1\strut}}
\newcommand{\CSLRightInline}[1]{\parbox[t]{\linewidth - \csllabelwidth}{\strut#1\strut}}

\usepackage{abstract}

\graphicspath{{figures/}}
\usepackage[margin=2.5cm]{geometry}
\ifLuaTeX
  \usepackage{selnolig}  
\fi
\usepackage{bookmark}
\IfFileExists{xurl.sty}{\usepackage{xurl}}{} 
\hypersetup{
  pdftitle={Structural controllability and management of cascading regime shifts},
  colorlinks=true,
  linkcolor={blue},
  filecolor={Maroon},
  citecolor={blue},
  urlcolor={blue},
  pdfcreator={LaTeX via pandoc}}

\title{Structural controllability and management of cascading regime
shifts}
\author{\small Juan C. Rocha\textsuperscript{1}, Anne-Sophie
Crépin\textsuperscript{2}\\
\small\\
\small \textsuperscript{1}Stockholm Resilience Centre, Stockholm
University\\
\small \textsuperscript{2}Beijer Institute of Ecological Economics, The
Royal Swedish Academy of Sciences}
\date{}

\begin{document}
\maketitle
\begin{abstract}
Abrupt transitions in ecosystems can be interconnected, raising
challenges for science and management in identifying sufficient
interventions to prevent them or recover from undesirable shifts. Here
we use principles of network controllability to explore how difficult it
is to manage coupled regime shifts. We find that coupled regime shifts
are easier to manage when they share drivers, but can become harder to
manage if new feedbacks are formed when coupled. Simulation experiments
showed that both network structure and coupling strength matter in our
ability to manage interconnected systems. This theoretical observation
calls for an empirical assessment of cascading regime shifts in
ecosystems and warns about our limited ability to control cascading
effects.
\end{abstract}

\section{Introduction}\label{introduction}

Regime shifts are large, abrupt and persistent changes in the function
and structure of systems\textsuperscript{1,2}. They have been documented
in a variety of social-ecological systems\textsuperscript{3,4}, as well
as climate\textsuperscript{5}, finance\textsuperscript{6}, or
health\textsuperscript{7}, to name a few examples. Changes in structure
and function of ecosystems can diminish the benefits people get from
nature such as food production, employment opportunities, climate
regulation, or water purification\textsuperscript{8}. For example, the
North Atlantic cod collapse represented millions of economic loss and
compromised more than 35000 jobs in 400 coastal
communities\textsuperscript{9}. The weakening or late arrival of the
Indian summer monsoon can compromise food production in one of the most
densely populated regions of the world. How management strategies avoid
regime shifts or recover natural ecosystems to desirable regimes are key
research questions and an active area of research.

But one theoretical fact that complicates their study and management is
that regime shifts can be interconnected\textsuperscript{10}. The
occurrence of one can impact the likelihood of another through
sequential tipping, like a domino effect\textsuperscript{10,11}; or
through two-way interactions where new feedbacks can amplify or dampen
the probability of tipping over\textsuperscript{10}. Conceptual work
have looked at these connections in the context of committed
risk\textsuperscript{5,12}, and recent modeling work has studied at
interactions between climate tipping elements to understand additional
challenges of prediction\textsuperscript{13}, or the role of minimal
network configurations\textsuperscript{14}. Besides climate, cascading
effects have also been studied in the context of engineered
systems\textsuperscript{15,16}, social and economic
systems\textsuperscript{19}, or diseases\textsuperscript{20}.

The possibility of cascading regime shifts raises a few additional
challenges for management. First, long distance coupling or
teleconnections\textsuperscript{21} between regime shifts implies that
optimal local or regional managerial practices might not be enough to
reach managerial objectives\textsuperscript{10,22}. This not only
illustrates the governance issue of with whom to coordinate for
successful management of local resources; it also implies a tragedy of
the commons with potential power asymmetries. Classical examples include
transnational resource systems where pollution or exploitation of a
resource on one side of the border impacts its quality on the other
side, or downstream\textsuperscript{23,24}. Second, and most
importantly, connected regime shifts pose the challenge of identifying
sufficient and appropriate points of intervention for managerial
actions. In an interconnected world, how can we identify leverage points
to manage regime shifts? Are there regime shifts -- or their
interactions - that are harder to manage?

\begin{figure*}[ht]
\centering
\includegraphics[width = 6.5in, height = 2.5in]{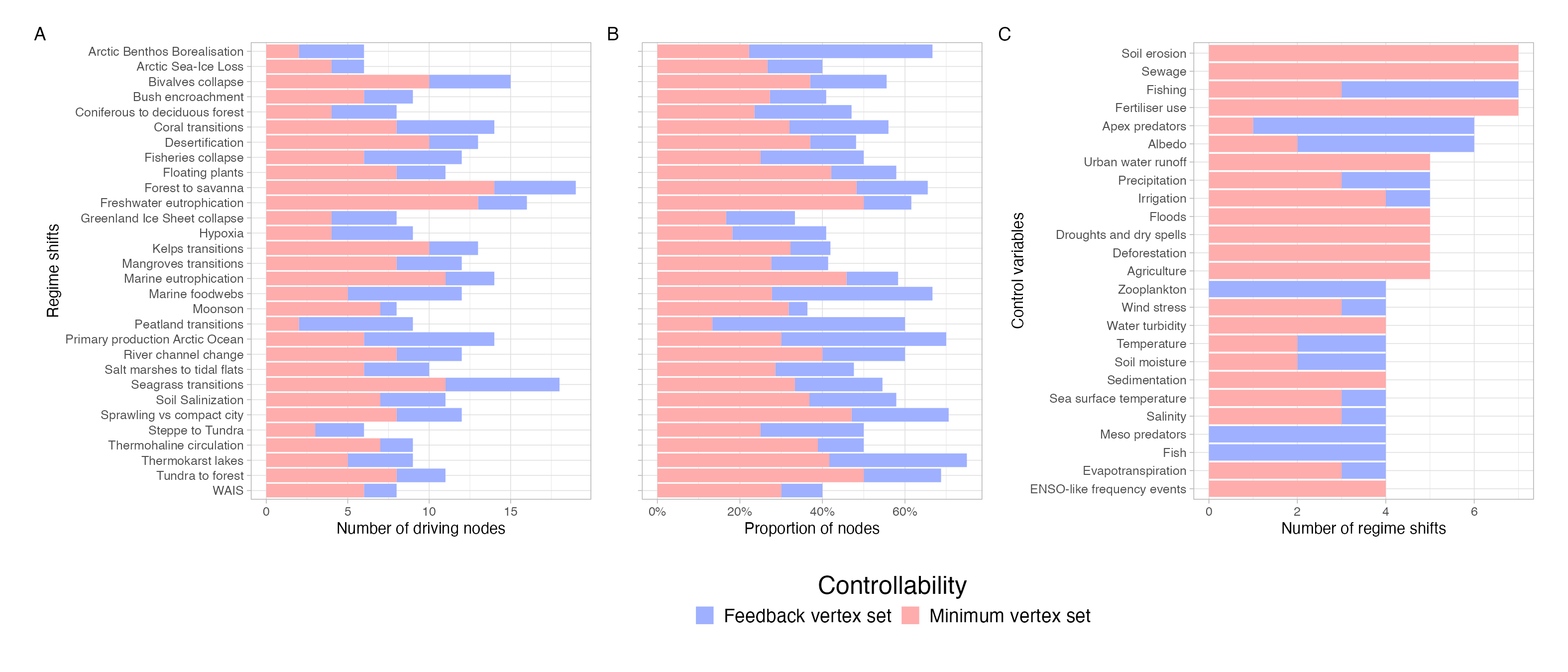}
\caption{\textbf{Structural controllability of individual regime shifts} The number of nodes of the control set per regime shift (A), and their proportion with respect to the full causal network (B). Top controlling variables ranked by the number of regime shifts where they are part of the control set.}
\label{fig:ind}
\end{figure*}

Here we answer these questions in the light of controllability of
complex systems. A system is controllable if one can drive it from any
initial state to any desired final state in finite
time\textsuperscript{25--27}. Hence, controllability quantifies our
ability to steer dynamical systems\textsuperscript{27}. While control
theory has been fairly well developed in the realm of linear systems,
many open problems remain for nonlinear ones\textsuperscript{26,28}.
Strict global controllability of nonlinear systems remain elusive, but
weaker notions of controllability can be
described\textsuperscript{26,29,30}. For example, open loop control has
been used to steer chaotic systems to desired trajectories in the phase
space. Here we focus on feedback vertex sets as a proxy of structural
controllability\textsuperscript{26,29} of regime shifts and their
interactions. We use network topology to gain insights on the
controllability of regime shift and the management challenges of their
cascading effects. We further test our intuitions with minimal models of
interconnected resources and pollution recipient systems.

\section{Methods}\label{methods}

We study structural controllability first in a static setting, and then
using minimal simulation experiments to approximate the dynamics of
interconnected systems. For the static setting we used causal diagrams
derived from the regime shifts database, an open online repository of
regime shift reviews\textsuperscript{4}. The database offers causal
graphs that summarize scientific hypotheses of how regime shifts work.
Each directed graph (N=30) can be interpreted as a network where a link
between two variables exists if a scientific paper has suggested a
causal link to the occurrence of the regime shift\textsuperscript{4}. We
use these networks to study structural controllability (Fig
\ref{fig:sm_clds}).

\textbf{Feedback control:} To steer a nonlinear system to a desired
attractor, one needs to find the feedback vertex set of the system. When
the system is described as a network, its state variables are
represented by nodes (or vertices), and the dependence between variables
as directed links (or arcs). A feedback in this context is a collection
of nodes and links that form a circular pathway. Feedbacks are often
responsible for the nonlinear dynamics of the system. If a network has
feedbacks, it is called a directed cyclic graph (DCG), and if it lacks
them a directed acyclic graph (DAG). The feedback vertex set is the set
of nodes that if deleted would render a DAG. The feedback arc set is the
set of links that if eliminated would render a DAG. Finding the feedback
vertex set or the arc set are equivalent problems\textsuperscript{31},
yet finding the minimum set is a nondeterministic polynomial-time hard
(NP-hard) problem. Once the system is reduced to a DAG, linear control
methods can be applied, such as finding unmatched nodes or the Kalman
rank condition for linear systems\textsuperscript{26,27}. Hence, the set
of nodes that need to be controlled to gain leverage on the behaviour of
a nonlinear system are the feedback vertex set plus the unmatched nodes
of the remaining linear system\textsuperscript{26,29}.

\textbf{Structural control on causal networks} We compute the proportion
of nodes that needs to be controlled for all causal networks from the
regime shifts database (Fig \ref{fig:sm_clds}), as well as their
pair-wise combinations. The pair-wise combinations have been used to
study plausible mechanisms that can couple different types of regime
shifts known as cascading effects\textsuperscript{10}. We then compare
how structural controllability changes if ecological regime shifts are
interconnected.

\begin{figure*}[ht]
\centering
\includegraphics[width = 6.5in, height = 2.5in]{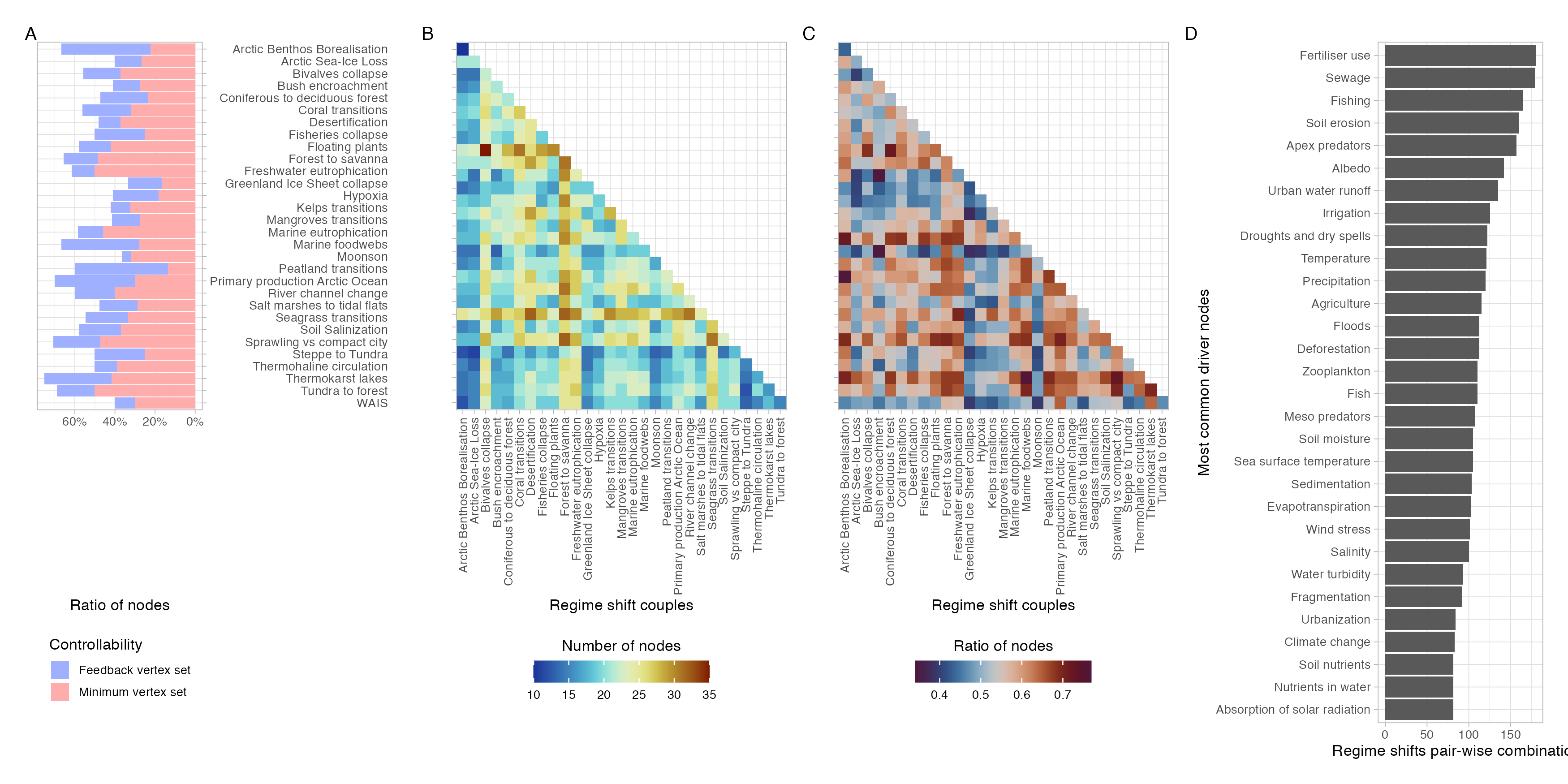}
\caption{\textbf{Structural controllability of coupled regime shifts} The proportion of nodes that need to be controlled to manage individual regime shifts (A) compared against the number of nodes (B) and ratio of nodes (C) that one needs to control to manage coupled regime shifts. Top controlling nodes are ranked by the number of pair-wise compbinations where they appear on the control set (D). For a sample of 30 regime shifts, there are 900 pair-wise combinations or 870 combinations of non-identical regime shifts. Thus, for example, fertilizers use and sewage appear in more than 175 of these combinations.}
\label{fig:coup}
\end{figure*}

\textbf{Dynamic control on minimal models} To further test our
intuitions, we compared the results from structural controllability with
modeling experiments in simplified resource and pollutant models. We
chose these models because they are easier to parametrise than the
qualitative networks offered by the database, yet they encapsulate well
understood and common dynamics underlying many real world regime shifts.
The pollution system is defined by:

\[
\frac{dx_i}{dt} = u_i-s_ix_i+v_i \frac{x_i^{\alpha_i}}{z_i^{\alpha_i}+x_i^{\alpha_i}}- \sum_{j\not=i}(\delta_{ij}x_i-\delta_{ji}x_j)A_{ij}
\]

where \(x_i\) is the level of pollutants of patch \(i\) (e.g.~a lake),
\(u_i\) denotes the level of pollutants from human activities, and
\(s_i\) the internal loss rate (e.g.~sedimentation). The term
\(v_i \frac{x_i^{\alpha_i}}{z_i^{\alpha_i}+x_i^{\alpha_i}}\) is a
pollutant release nonlinear function where \(v_i\) is the maximum level
of internal pollutants, \(\alpha_i\) indicates the sharpness of the
function, and \(z_i\) is a threshold level at which the system flips
from low to high nutrient regimes. Connectivity across patches is
represented with the term
\(\sum_{j\not=i}(\delta_{ij}x_i-\delta_{ji}x_j)A_{ij}\) where
\(\delta_{ij}\) is the diffusion coefficient from patch
\(i \rightarrow j\), given that \(\delta_{ij}\) and \(\delta_{ji}\) are
not necessarily the same; while \(A_{ij}\) is the adjacency matrix.
Similarly, the resource system is defined by:

\[
\frac{dy_i}{dt} = r_iy_i\left(1-\frac{y_i}{k_i}\right)-c_i \frac{y_i^{\beta_i}}{q_i^{\beta_i}+y_i^{\beta_i}} + \sum_{j\not=i}(\delta_{ij}y_j- \delta_{ji}y_i)A_{ij}
\]

where \(y_i\) is the state of the population \(i\), \(r_i\) is the
intrinsic growth rate, and \(k_i\) its carrying capacity. The term
\(c_i \frac{y_i^{\beta_i}}{q_i^{\beta_i}+y_i^{\beta_i}}\) is a Holling
type III predation function where \(c_i\) represent refuges from
predators in patch \(i\), \(q_i\) is the half saturation biomass for
predators (e.g.~a threshold for high to low abundance), and \(\beta_i\)
is the curvature parameter that defines the abruptness of the
shift\textsuperscript{32}. The last term indicates mobility of the
resource between patches at a diffusion rate
\(\delta_{ij} \neq \delta_{ji}\), and an adjacency matrix \(A_{ij}\)
that takes 1 if patches \(i\) and \(j\) are connected, or zero
otherwise. For both systems the diffusion coefficient (\(\delta_{i,j}\))
is normalized by the out degree, or the number of outgoing connections
from each system. This is to ensure that the amount of pollutants or
resources flowing from one system to another never exceeds the amount
available at run time.

Here we concentrate our efforts in exploring numerical simulations of
medium size interconnected systems (\(N < 10^2\)), with potentially
multiple feedbacks. Analytical solutions for the low dimensional
diffusion problem of our pollution and resource systems can be found
in\textsuperscript{33}. For each of the dynamical systems (pollution,
resource) we generated networks varying its size
(\(N:\{25, 50, 75, 100\}\)); the network generative process which could
be random (Erdös-Rényi model), preferential attachment (Barabási-Albert
model), or small-world (Watts-Strogatz model); the network density
(\(d:\{0.05, 0.1, 0.3, 0.5\}\)); and the coupling strength
(\(\delta_{i,j}:\{low = 0.01:0.05, high = 0.25:0.5\}\)). Thus, we have
96 different experimental conditions for which we generated 100
replicates resulting in 9600 simulation experiments for each dynamical
system. Each model was run for 200 time steps and integration steps of
0.001 \(dt\).

For each experiment we identified the feedback vertex set and the
minimum controlling set of the network. Then we manipulated the driver
at run time to decrease nutrients in the pollution system (\(u\)
decreased from 5 to -5 allowing for management that can extract
nutrients from the system), or decrease harvest in the resource system
(\(c\) decreased from 10 to 0) on the controlling set only. We then
assess if management of the control set successfully recovers the
networked systems, the proportion of nodes recovered, or time to
recovery when comparing strong versus weak coupling (\(\delta\)). The
computer code used for our simulations as well as the curated networks
from the regime shifts database are available at:
\url{https://github.com/juanrocha/imperio}

\section{Results}\label{results}

Individual regime shifts cannot be controlled or managed by only
targeting variables that have been coded as drivers in the regime shifts
database. Drivers there are defined as variables outside feedbacks that
influence the dynamics of the system, but that are not in turn
influenced by the system itself\textsuperscript{4}; which inherits the
logic first proposed by the Millennium Ecosystem
Assessment\textsuperscript{34}. As recent advances in network
controllability show\textsuperscript{26,29}, to steer the system towards
a desired state, an ideal manager needs to intervene not only on drivers
but also on some of the variables involved in feedback dynamics. In all
the individual regime shifts analyzed, an ideal manager needs to account
for the minimum control set of the DAG and the feedback vertex set (Fig
\ref{fig:ind}). By number of driving nodes, regime shifts such as forest
to savanna or sea grass transitions require higher efforts; but by
proportion of nodes: thermokarst lakes, primary productivity in the
Arctic ocean, tundra to forest, or marine food webs are among the regime
shifts where \textgreater60\% of nodes need some intervention. The top
controlling variables are, as expected, related to climate change,
biodiversity loss, or food production (Fig \ref{fig:ind}C). We observe
that trophic groups such as zooplankton, meso predators, or top
predators are often part of the feedback vertex set, while fertilizers
use, sewage or soil erosion are always part of the minimum control set
of the DAG.

Coupled regime shifts are harder to manage because the ideal manager
needs to influence more variables. However, the number and proportion of
nodes to be controlled varies depending on the coupling (Fig
\ref{fig:coup}). On average, coupled regime shifts are harder to manage
than individual counterparts (Fig \ref{fig:coup}). For example,
preventing the West Antarctica Ice Sheet collapse (WAIS) implies being
able to intervene on a minimum of 40\% of the variables describing
processes underlying the regime shift. But when combined with other
regime shifts, the percentage increases to 49\% on average, and a
maximum of 64\%. Similarly, managing marine food webs regime shifts
requires intervening on \textasciitilde66\% of the variables describing
the individual shift; but when coupled it varies from as low as 53\%
with kelp transitions to as high as 72\% with Arctic benthos
borealization. This is because sharing common variables on the minimum
driving set reduces the proportion of nodes to control, while new
feedbacks formed when coupling the network, increase the feedback vertex
set. The most common variables to manage coupled regime shifts are
fertiliser use, sewage, fishing and soil erosion. Climate related
variables are important but down the rank in the list, such as droughts,
temperature, and precipitation. Many variables that have not been
previously reported in similar studies about key drivers of regime
shifts are reported here because they are part of the feedback vertex
set, such as management of apex predators, zooplankton, or fish
populations. This however aligns with management strategies that attempt
to manipulate feedback strength to recover ecosystems from regime
shifts, for example fish population manipulations to recover from
eutrophication\textsuperscript{35}.

\begin{figure}[ht]
\centering
\includegraphics[width = 3in, height = 4.5in]{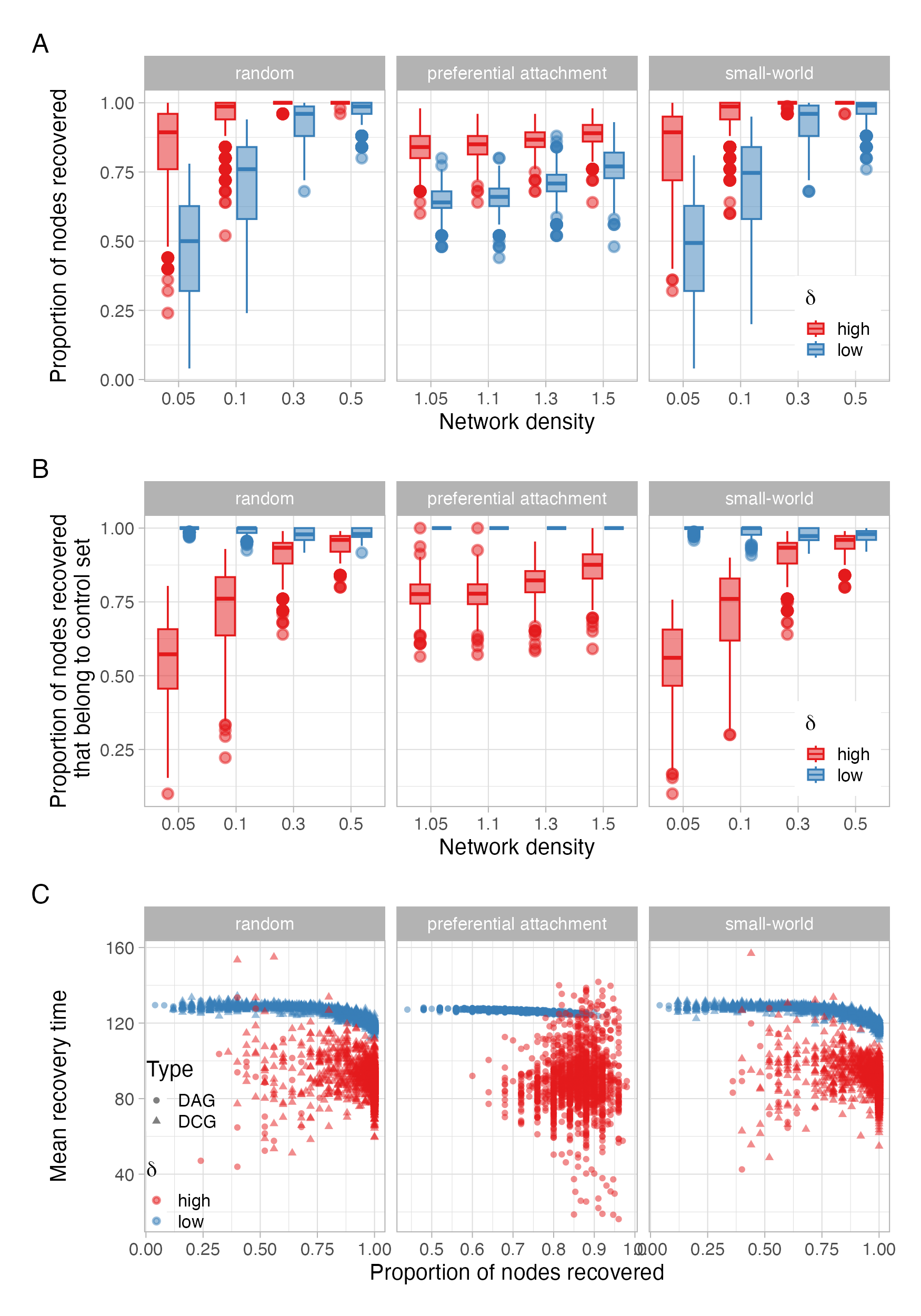}
\caption{\textbf{Pollution system simulations} The proportion of nodes recovered by network density for different types of network architecture and coupling strength (A). The proportion of nodes recovered that belongs to the control set (B), and the mean recovery time by proportion of nodes recovered (C). Boxplots and scaterplots summarize N = 9600 networks (100 replicates * 96 experimental settings).}
\label{fig:pollution}
\end{figure}

These results suggest that network structure can have implications for
management. If two regime shifts are solely coupled by sharing drivers,
then controlling the common drivers will help the ideal manager to
tackle both regime shift problems. But if the coupling of regime shifts
generates new feedbacks, the minimum number of processes that need to be
controlled increases, making it more difficult to manage the coupled
system than the independent ones. To test that intuition, we developed
simple toy models of resource and pollutant systems where we varied
network structure. The models were not as complex as the causal graphs
previously studied, but they still capture the essential dynamics. Both
typeS of systems in isolation can be controlled by one parameter, the
addition or removal of pollutants \(u_i\), and the availability of
refuges from predators \(c_i\) respectively. Thus when coupled, any
gains or losses on the ability to manage the regime shifts is due to
network structure alone (the coupling matrix \(A_{i,j}\)).

\begin{figure}[ht]
\centering
\includegraphics[width = 3in, height = 4.5in]{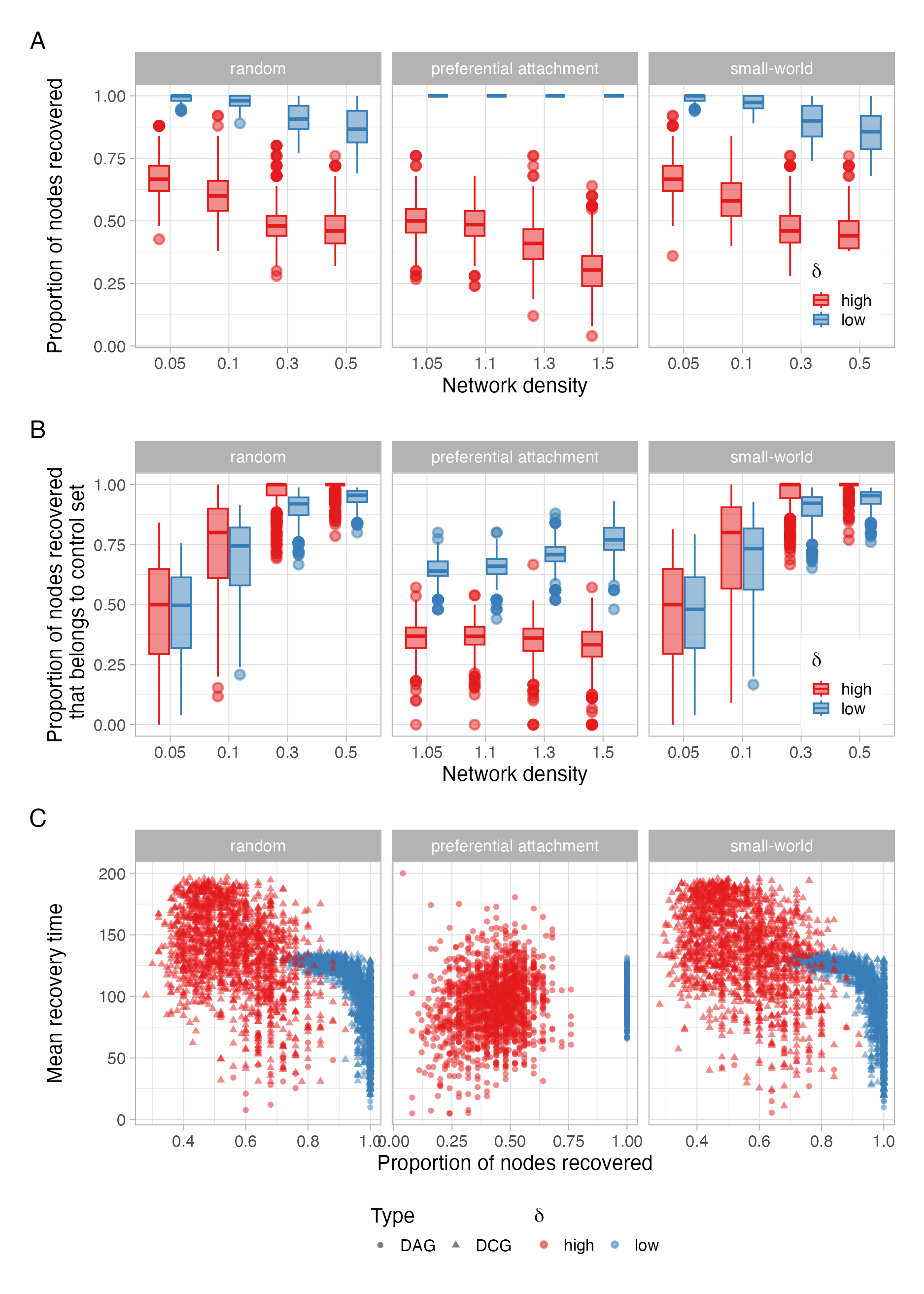}
\caption{\textbf{Resource system simulations} The proportion of nodes recovered by network density for different types of network architecture and coupling strength (A). The proportion of nodes recovered that belongs to the control set (B), and the mean recovery time by proportion of nodes recovered (C). Boxplots and scaterplots summarize N = 9600 networks (100 replicates * 96 experimental settings).}
\label{fig:resource}
\end{figure}

In the pollution system, the higher the network density the higher the
proportion of nodes recovered in our simulations (Fig
\ref{fig:pollution}A). This effect is accentuated when the coupling
strength \(\delta\) is high. However, the proportion of nodes recovered
that belong to the control set is higher when coupling strength is low
(Fig \ref{fig:pollution}B) for all network types tested. This means that
coupling strength increases the probability that a larger proportion of
the system is recovered by managerial actions on the control set, but
the control set in turn becomes vulnerable through incoming connections
that impede complete recovery. We also observe that mean recovery time
is on average faster with high coupling strength, but in networks whose
generative process is preferential attachment (scale-free like degree
distributions) and strong presence of dyadic acyclic graphs (DAG),
recovery can sometimes be slower under high coupling (Fig
\ref{fig:pollution}C). Regardless of network architecture, the lower
bound for proportion of node recovery is as low as 60\% for preferential
attachment networks, around 30\% for random and small-world networks
under high coupling strength, and almost none for low coupling.

We find contrasting results for the resource simulations. First, the
proportion of nodes recovered declines on average with increasing
network density, although for the case of preferential attachment, all
nodes were recovered regardless of network density in the low coupling
strength scenario (Fig \ref{fig:resource}A). Second, the proportion of
nodes recovered that belong to the control set increases with network
density for the cases of random and small-world networks, but there are
no differences between the high and low coupling strength (Fig
\ref{fig:resource}B). In the case of preferential attachment, the
proportion of recovered nodes that belong to the control set is higher
in networks under the low coupling strength. High coupling strength
reduces the chances of recovery for the control set. While most
simulated systems under low coupling strength recovered by time step
\textasciitilde130, recovery time varied much more in networks with high
coupling strength (Fig \ref{fig:resource}C).

\section{Discussion}\label{discussion}

Our simulation results confirm that network structure matters when
trying to manage coupled regime shifts. The type of network architecture
(random, preferential attachment, small-world) changes the likelihood of
recovering the full system, as well as the coupling strength between
systems. Surprisingly, even under the same network configurations
(N=9600) the type of dynamics occurring in the networks (pollutant
versus resource equations) substantially influenced the qualitative
patterns identified for the full ensemble (Figs \ref{fig:pollution},
\ref{fig:resource}).

In many of our experimental settings the system only recovered partially
despite having full information of the control set at the start of the
simulation. One plausible explanation is that the system required longer
time to recover. A more likely explanation however is that our network
setting allowed for changes in the amount and direction of flow of
nutrients and pollutants. In both systems the diffusion term depended on
the difference in concentration between patches
(\(\sum_{j\not=i}(\delta_{ij}y_j-\delta_{ji}y_i)A_{i,j}\)). While the
adjacency matrix \(A_{i,j}\) and the diffusion coefficients
\(\delta{i,j}\) were constant through the simulation, the direction of
flow could change if the difference \(\delta_{ij}y_j-\delta_{ji}y_i\)
was positive or negative. In practice, this means that the feedback
structure and thus the control set can change over time. Another
plausible explanation is that the adjacency matrix (\(A_{ij}\)) is
non-normal. Networks are non-normal if \(AA^T \ne A^TA\), which is the
case in our directed graphs, although matrix asymmetry is not sufficient
for strong non-normality\textsuperscript{36}. Strong non-normality
emerges when the network is directed (asymmetrical), has low
reciprocity, lacks feedback, and has hierarchical
organization\textsuperscript{36}. While our networks have feedbacks, we
cannot discard the possibility that the lack of recovery is due to
non-normal induced features on the dynamics. The non-normality of
networks increases their return time to equilibrium, thus reducing their
resilience\textsuperscript{37}.

Knowing the control set at one moment in time is insufficient to gain
full control, and thus recovery, of the connected system. Deriving the
control set at the beginning of the simulation resulted in insufficient
efforts to successfully stir the system towards full recovery (Figs
\ref{fig:pollution}, \ref{fig:resource}). In practice, this is likely to
be the case if one is managing a meta-population where organisms are
free to move from patch to another and likely to go to places where they
find better resources, which change over time. It is less likely to
occur in settings where the flow of resources or pollutants is bounded
by a physical parameter. For example, water between lakes flows
according to gravity, so nutrients are unlikely to be transported
against it. While our assumption can be unrealistic for certain cases,
it opens the case when the control set changes over time. In most real
settings we do not even know the real empirical structure of the network
--how regime shifts are connected\textsuperscript{10}. Our results
suggest that even if we knew the structure at a certain point in time,
successfully steering the networked system requires monitoring programs
that enable learning the structure, direction and coupling strength of
the system in real time.

This dynamical feature is unlikely to affect our structural
controllability results derived from causal networks (Fig
\ref{fig:coup}), because the direction of causality is unlikely to
change. However, in a dynamical setting the feedback strength or the
coupling of different variables could change, akin to \(\delta{i,j}\) in
our simulations. Thus to apply efficient managerial practices that avoid
tipping cascades, one could gain further insights on the priority of
variables for management, or sequence of actions required. Our results
showed that while climate is important, other coupling mechanisms can be
more relevant in the sense that they appear more in pair-wise
combinations (Fig \ref{fig:coup}). Our results also confirm that dealing
with climate related variables alone won't be enough to prevent tipping
cascades\textsuperscript{10}, after all, these variables are a small
share of the control set.

Our modelling framework opens potential avenues for future research.
Thus far our assumption is that an ideal manager can intervene in all
subsystems of the controlling set simultaneously. Relaxing thus
assumption would question whether it is possible to recover the system
with asynchronous control. We have not yet considered the cost of
interventions in the system, nor time bounds. With limited budget or
time to achieve the controllability goals, what would be an optimal
strategy? or a sequence of strategies? Another interesting avenue for
future research is a case with multiple managers who have to coordinate
to avoid the tipping cascade, adding a game theoretic layer to the
problem. We believe that relaxing any of our simplifying assumptions to
make the problem more realistic would inevitably make it harder to
control.

Our results show that interconnected systems put certain subsystems and
their managers in a privileged position to steer the system, for good
and bad. Given these power asymmetries, is coordination and cooperation
likely to emerge to avoid tipping cascades? The question is imperative
since both the climate and biodiversity crises underlie a similar
networked structure to the framework here proposed, where a few actors
have disproportional agency to influence the overall dynamics of the
system. Moreover, our results show that when systems are interconnected,
there is no guarantee that one can recover the full system. This implies
that current climate narratives that rely on the potential recovery from
overshoot scenarios\textsuperscript{38} should be taken with skepticism,
because the probability of recovery seen in simple one-dimensional
models does not necessarily hold in higher dimensions. Here we show that
the ability to control is reduced. In fact, if one is to believe that
tipping points can be interconnected\textsuperscript{5,10,11}, each
connections becomes a co-dimension (a driver) of other tipping points,
so the problem is never one-dimensional.

\section{Conclusion}\label{conclusion}

By applying principles of structural controllability, we show here that
connected regime shifts are more difficult to manage and recover than
independent systems. Sharing common drivers can perhaps simplify the
problem, but the emergence of feedbacks increases the number of
variables that need interventions. Interconnected systems come with
additional dynamic challenges for network control and management: being
able to observe and measure how the control set changes over time. In
more realistic settings, the network structure imposes additional
challenges for the emergence of cooperation, power asymmetries, the
cost, timing, and sequence of managerial actions. These questions remain
open areas of research. Yet, tackling the climate or biodiversity crises
requires learning such networks from empirical data. Our theoretical
exercise shows that management practices that can work on one
dimensional systems such as overshooting scenarios for climate may not
necessarily succeed in high dimensional systems. We appeal for the
precautionary principle when dealing with interconnected crises.

\section{Acknowledgements}\label{acknowledgements}

We acknowledge support from VR grant 2022-04122, Formas grants
2015-00731, 2020-00454 and 2019-02316, the latter through the support of
the Belmont Forum.

\section{References}\label{references}

\small

\phantomsection\label{refs}

\onecolumn

\section{Supplementary Material}\label{sec:SM}

\renewcommand\thefigure{S\arabic{figure}}
\renewcommand\thetable{S\arabic{table}}
\setcounter{table}{0}
\setcounter{figure}{0}
\begin{figure}[ht]
\centering
\includegraphics[width = 6in, height = 6in]{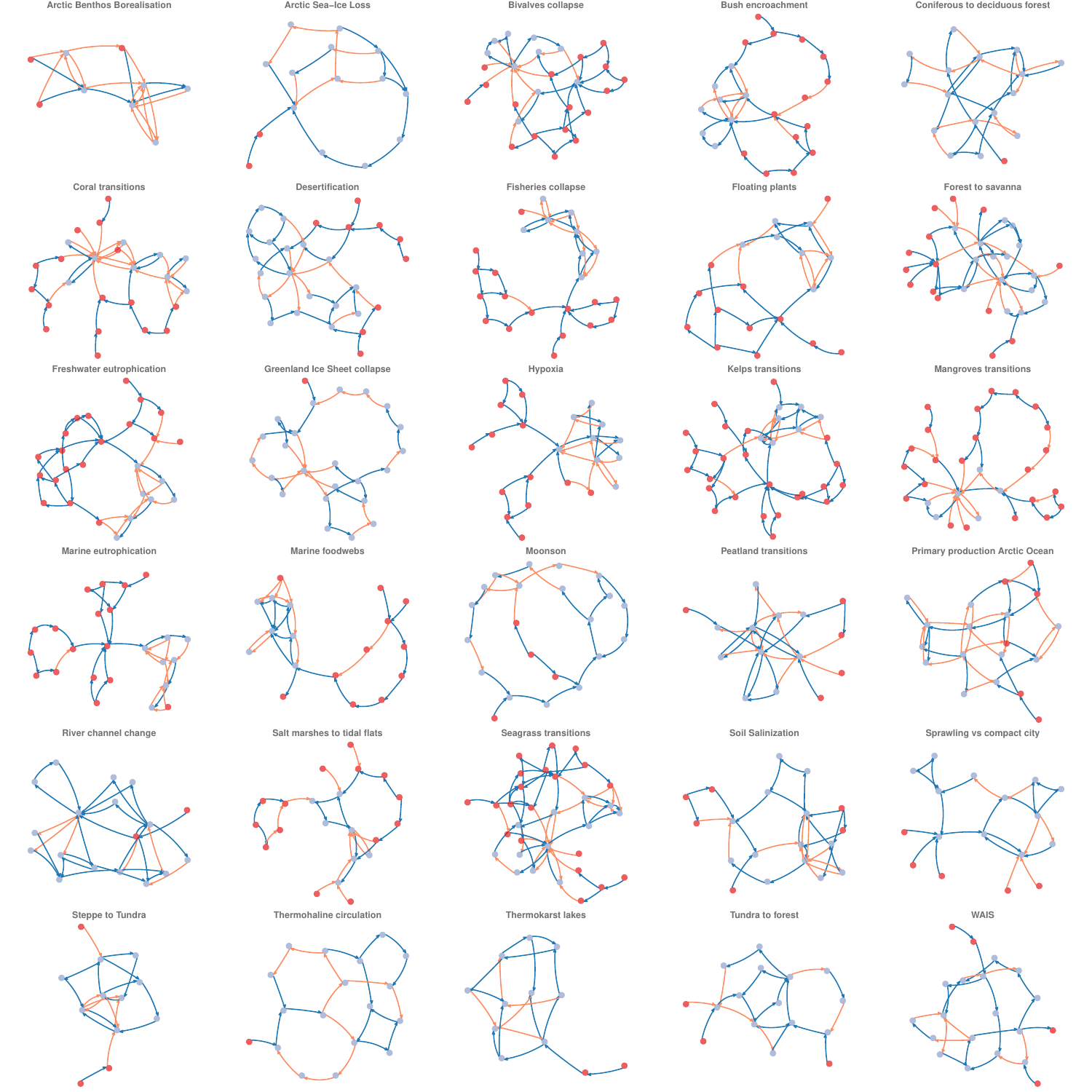}
\caption{\textbf{Causal networks} The regime shifts database (www.regimeshifts.org) reports causal hypothesis of how regime shifts work synthesized in causal loop diagrams, where variables are connected by arrows if a causal mechanisms is reported in scientific literature. Variables in red are drivers (outside feedbacks), in grey are variables that belong to feedback mechanisms, links in blue are positive and orange negative. We used the version of the database reported in ref 10.}
\label{fig:sm_clds}
\end{figure}

\end{document}